\documentclass[a4paper,fleqn]{cas-dc}

\usepackage[english]{babel}
\usepackage[numbers,sort&compress]{natbib}
\usepackage{graphicx}
\usepackage{graphics}
\usepackage{braket}
\usepackage{bbold}
\usepackage{amssymb}
\usepackage{amsmath}
\usepackage{nicefrac}
\usepackage{dcolumn}
\usepackage{bm}
\usepackage{slashed}
\usepackage{datetime}
\usepackage{mciteplus}
\usepackage{multirow}
\usepackage{bbold}
\usepackage{color}
\usepackage{xcolor}
\usepackage{float}
\usepackage[utf8]{inputenc}
\usepackage[normalem]{ulem}

\def\tsc#1{\csdef{#1}{\textsc{\lowercase{#1}}\xspace}}
\tsc{WGM}
\tsc{QE}

\newcommand{\drv}{{\rm d}}

\newcommand{\LQCD}{\Lambda_{\rm QCD}}
\newcommand{\MSb}{\overline{\rm MS}}

\newcommand{\DY}{\Delta Y}

\newcommand{{\Jethad}}{\textsc{Jethad}}
\newcommand{{\Powheg}}{\textsc{Powheg}}

\begin{document}
\let\WriteBookmarks\relax
\def\floatpagepagefraction{1}
\def\textpagefraction{.001}

\shorttitle{The high-energy QCD dynamics from Higgs-plus-jet correlations at the FCC}    

\shortauthors{Celiberto, Francesco Giovanni}
\shortauthors{Papa, Alessandro} 

\title []{\Huge The high-energy QCD dynamics from \\ Higgs-plus-jet correlations at the FCC}  

\author[1,2,3,4]{Francesco Giovanni Celiberto}[orcid=0000-0003-3299-2203]
\cormark[1]
\author[5,6]{Alessandro Papa}[orcid=0000-0001-8984-3036]


\ead{francesco.celiberto@uah.es}
\ead{alessandro.papa@fis.unical.it}


\affiliation[1]{organization={European Centre for Theoretical Studies in Nuclear Physics and Related Areas (ECT*)},
            addressline={Strada delle Tabarelle 286}, 
            city={Villazzano},
            postcode={I-38123}, 
            state={Trento},
            country={Italy}}

\affiliation[2]{organization={Fondazione Bruno Kessler (FBK)},
            addressline={Via Sommarive 18}, 
            city={Povo},
            postcode={I-38123}, 
            state={Trento},
            country={Italy}}

\affiliation[3]{organization={INFN-TIFPA Trento Institute of Fundamental Physics and Applications},
            addressline={Via Sommarive 14}, 
            city={Povo},
            postcode={I-38123}, 
            state={Trento},
            country={Italy}}

\affiliation[4]{organization={Universidad de Alcal\'a (UAH)},
            addressline={Plaza San Diego, s/n}, 
            city={Alca\'a de Henares},
            postcode={E-28801}, 
            state={Madrid},
            country={Spain}}

\affiliation[5]{organization={Dipartimento di Fisica, Universit\`a della Calabria},
            addressline={Ponte Pietro Bucci, Cubo 31C}, 
            city={Arcavacata di Rende},
            postcode={I-87036}, 
            state={Cosenza},
            country={Italy}}

\affiliation[6]{organization={INFN, Gruppo Collegato di Cosenza},
            addressline={Ponte Pietro Bucci, Cubo 31C}, 
            city={Arcavacata di Rende},
            postcode={I-87036}, 
            state={Cosenza},
            country={Italy}}

\cortext[1]{Presenter at the \textbf{FCC Week 2022}, 30 May - 8 June 2022, Campus des Cordeliers - Sorbonne Universit\'e.}



\begin{abstract}
Recent analyses on high-energy inclusive Higgs-boson rates in proton collisions \emph{via} the gluon-fusion channel, matched with the state of-the-art fixed-order N$^3$LO accuracy, have shown that the impact of high-energy resummation corrections reaches 10\% at the FCC nominal energies.
This supports the statement that electroweak physics at 100 TeV is expected to receive relevant contributions from small-$x$ physics.
In this preliminary study we present novel predictions for azimuthal-angle and rapidity distributions sensitive to the inclusive emission of a Higgs boson in association with a light-flavored jet in proton collisions, calculated within the NLL accuracy of the energy-logarithmic resummation.
We highlight how high-energy signals for this process are already present and visible at current LHC energies, and they are also sizable at the FCC ones.
We come out with the message that the improvement of fixed-order calculations on Higgs-sensitive QCD distributions is a core ingredient to reach the precision level in the description of observables relevant for the Higgs physics at the FCC.
\end{abstract}




\maketitle

\section{Introductory remarks}
\label{sec:introduction}

A new era for high-energy physics started with the discovery of the Higgs boson at the Large Hadron Collider (LHC)~\cite{ATLAS:2012yve,CMS:2012qbp}.
Since then, a joint effort has been made from the experimental and the theoretical sides both to provide us with accurate benchmarks of the Standard Model and to seek for distinctive signals of New Physics.
From a formal viewpoint, higher-order computations of Higgs production in perturbative Quantum Chromodynamics (QCD) are needed to make stringent tests of both the gluon and the vector-boson fusion subreactions.
In the case of gluon fusion, the presence of top-quark loops even at leading order (LO)~\cite{Georgi:1977gs}, makes performing those calculations not a straightforward task.
An effective description, whose validity holds in the large top-mass limit~\cite{Wilczek:1977zn}, represented a first and common basis for pioneering next-to-leading order (NLO) QCD analyses of the gluon-fusion channel~\cite{Dawson:1990zj,Djouadi:1991tka}, while the next-to-NLO (NNLO)~\cite{Harlander:2002wh,Anastasiou:2002yz,Ravindran:2003um} and the next-to-next-to-NLO (N$^3$LO) accuracy level~\cite{Anastasiou:2014lda,Anastasiou:2015vya,Anastasiou:2016cez,Mistlberger:2018etf} were subsequently reached.
The weight of finite top-mass effects at NLO, NNLO, and N$^3$LO was assessed in Refs.~\cite{Spira:1995rr,Harlander:2009my,Harlander:2009mq,Harlander:2009bw,Pak:2009bx,Pak:2009dg,Davies:2019wmk}.
Then, inclusive distributions for the emission of Higgs-plus-jet systems were investigated in the NLO~\cite{Bonciani:2016qxi,Jones:2018hbb,Bonciani:2022jmb} and NNLO~\cite{Boughezal:2013uia,Chen:2014gva,Boughezal:2015dra,Boughezal:2015aha} QCD.

The aforementioned computations are relevant to probe the well-established \emph{collinear factorization}~\cite{Collins:1989gx}, where cross sections are written as a one-dimensional convolution of on-shell coefficient functions with collinear parton density functions (PDFs) and, in case of final-state identified hadrons, also with collinear fragmentation functions (FFs).
While the collinear formalism has achieved striking successes in the study of several high-energy reactions at hadron and lepton-hadron colliding machines, there are kinematic sectors where one or more types of logarithms, genuinely missed by the standard collinear formalism, are enhanced. These logarithms can be large enough to prevent the convergence of the perturbative series.
Thus, adequate all-order techniques, known as \emph{resummations}, must be used to catch these logarithmic corrections.

One of these sectors is the so-called \emph{semi-hard} regime \cite{Gribov:1983ivg,Celiberto:2017ius}, where a 
scale order, $\sqrt{s} \gg \{Q\} \gg \LQCD$, ($s$ is the center-of-mass energy squared, $\{Q\}$ stands for one or a set of hard scales typical of the reaction, and $\LQCD$ is the QCD scale) stringently holds.
Here, large-energy logarithms, $\ln(s/Q^2)$, become relevant.
The most adequate tool to perform the high-energy resummation is the Balitsky--Fadin--Kuraev--Lipatov (BFKL) formalism~\cite{Fadin:1975cb,Kuraev:1976ge,Kuraev:1977fs,Balitsky:1978ic},  
defined both in the leading-logarithmic (LL) and in the next-to-leading logarithmic (NLL) approximation.
A BFKL-resum\-med cross section is cast as a high-energy convolution between a universal Green's function, whose kernel is known up the NLO~\cite{Fadin:1998py,Ciafaloni:1998gs,Fadin:1998jv,Fadin:2000yp,Fadin:2004zq,Fadin:2023roz}, and process-dependent singly off-shell coefficient functions, also known as forward \emph{impact factors}.
Remarkably, the BFKL approach has offered us the possibility to access the proton content at small-$x$ \emph{via} several different types of (un)integrated distributions~\cite{Ball:2017otu,Abdolmaleki:2018jln,Bonvini:2019wxf,Bacchetta:2020vty,Celiberto:2021zww,Hentschinski:2012kr,Bolognino:2018rhb,Bolognino:2021niq,Celiberto:2019slj,Hentschinski:2021lsh,Nefedov:2021vvy}.

Promising channels to probe the semi-hard QCD domain at hadron colliders are inclusive hadroproductions of two objects emitted with large transverse masses and well separated in rapidity.
The last condition is required to heighten the weight of secondary-gluon radiation featuring a strong ordering in rapidity. From a mathematical point of view, this translates into a rise of the aforementioned energy logarithms, which must be resummed to all orders by BFKL.
Cross sections for these reactions can be written by the hands of a \emph{hybrid high-energy and collinear factorization}, where the standard description is supplemented by the $t\text{-}$channel resummation of energy logarithms.
Since reaching the highest possible logarithmic accuracy, namely the NLL level, requires to calculate both the BFKL Green's function and the off-shell coefficient functions at NLO in perturbative QCD, we refer to our formalism as NLL/NLO$^{(*,+)}$ hybrid factorization~\cite{Colferai:2010wu,Celiberto:2020tmb,Bolognino:2021mrc,Celiberto:2022rfj,Deak:2009xt,Deak:2018obv,vanHameren:2022mtk}.
The possible presence of a `$*$' superscript indicates that, in our treatment for a given observable, some NLO terms are missing.
This generally happens when not all the NLO corrections of a considered coefficient function have been calculated yet.
Conversely, a `$+$' superscript indicates that some extra next-to-NLL contributions have been included in an effective way.

NLL/NLO$^{(+)}$ studies of two jets widely separated in rapidity~\cite{Colferai:2010wu,Ducloue:2013hia,Ducloue:2013bva,Caporale:2014gpa,Ducloue:2015jba,Celiberto:2015yba,Celiberto:2015mpa,Celiberto:2016ygs,Celiberto:2022gji} (Mueller--Navelet channel~\cite{Mueller:1986ey}) were compared with LHC data at $\sqrt{s} = 7\mbox{ TeV}$~\cite{Khachatryan:2016udy}.
NLL/NLO$^{(*)}$ analyses on three-jet tags~\cite{Caporale:2016zkc}, light-flavored hadron detections~\cite{Celiberto:2016hae,Celiberto:2017ptm,Bolognino:2018oth,Celiberto:2020rxb,Celiberto:2022kxx}, forward Drell--Yan emissions~\cite{Celiberto:2018muu,Golec-Biernat:2018kem}, and heavy-flavored jet tags~\cite{Bolognino:2019ccd,Bolognino:2019yls} soon followed.
Studies on heavy-hadron production \emph{via} variable-flavor number-scheme (VFNS) FFs~\cite{Mele:1990cw,Cacciari:1993mq} have provided evidence that that the peculiar behavior of the gluon fragmentation channels permits to \emph{naturally stabilize} the BFKL series~\cite{Celiberto:2022grc}. This makes it reliable to employ the NLL/NLO$^{(+)}$ hybrid factorization at the \emph{natural scales} provided by process kinematics~\cite{Celiberto:2021dzy,Celiberto:2021fdp,Celiberto:2022dyf,Celiberto:2023fzz,Celiberto:2022keu}.
Such a result cannot be achieved when light jets and/or hadrons are detected~\cite{Celiberto:2020wpk}, due to the rise of genuine NLL instabilities as well as \emph{threshold} contaminations.

Another promising channel to hunt for the manifestation of stabilization effects is the semi-inclusive Higgs-plus-jet production at hadron colliders. 
Here, precision studies of the transverse-momentum spectrum rely upon a doubly differential resummation both in the Higgs $p_T$ and in the leading-jet $p_T$-veto~\cite{Monni:2019yyr}.
When, however, the rapidity interval between the two final-state objects increases, the weight of energy logarithms grows.
Therefore, the high-energy resummation, encoded in our hybrid factorization, can serve as a useful tool to possibly improve the description of Higgs-plus-jet differential distributions.
On the one hand, recent analyses on the high-energy, rapidity-inclusive Higgs-boson production in gluon fusion at the LHC, matched with the state of-the-art fixed-order N$^3$LO accuracy~\cite{Bonvini:2018ixe,Bonvini:2018iwt}, have shown that the impact of high-energy resummation corrections reaches 10\% at the Future Circular Collider (FCC) nominal energy of $\sqrt{s} = 100$~TeV~\cite{FCC:2018vvp,FCC:2018byv}.
On the other hand, Higgs-plus-jet distributions might be sensitive to the high-energy resummation at lower energies, such as the current LHC ones, provided that the rapidity interval between the Higgs and the jet is sufficiently large~\cite{DelDuca:1993ga,DelDuca:2003ba,Celiberto:2020tmb,Andersen:2022zte}.

In this short document we present preliminary results on an ongoing extension of the study of Ref.~\cite{Celiberto:2020tmb}, focused on analyzing the behavior of Higgs-plus-jet differential distributions within a NLL/NLO$^*$ hybrid treatment and at 100~TeV~FCC.

\section{Higgs-plus-jet production in hybrid factorization at NLL/NLO*}
\label{sec:theory}

\begin{figure}[t]
\centering
\includegraphics[width=0.40\textwidth]{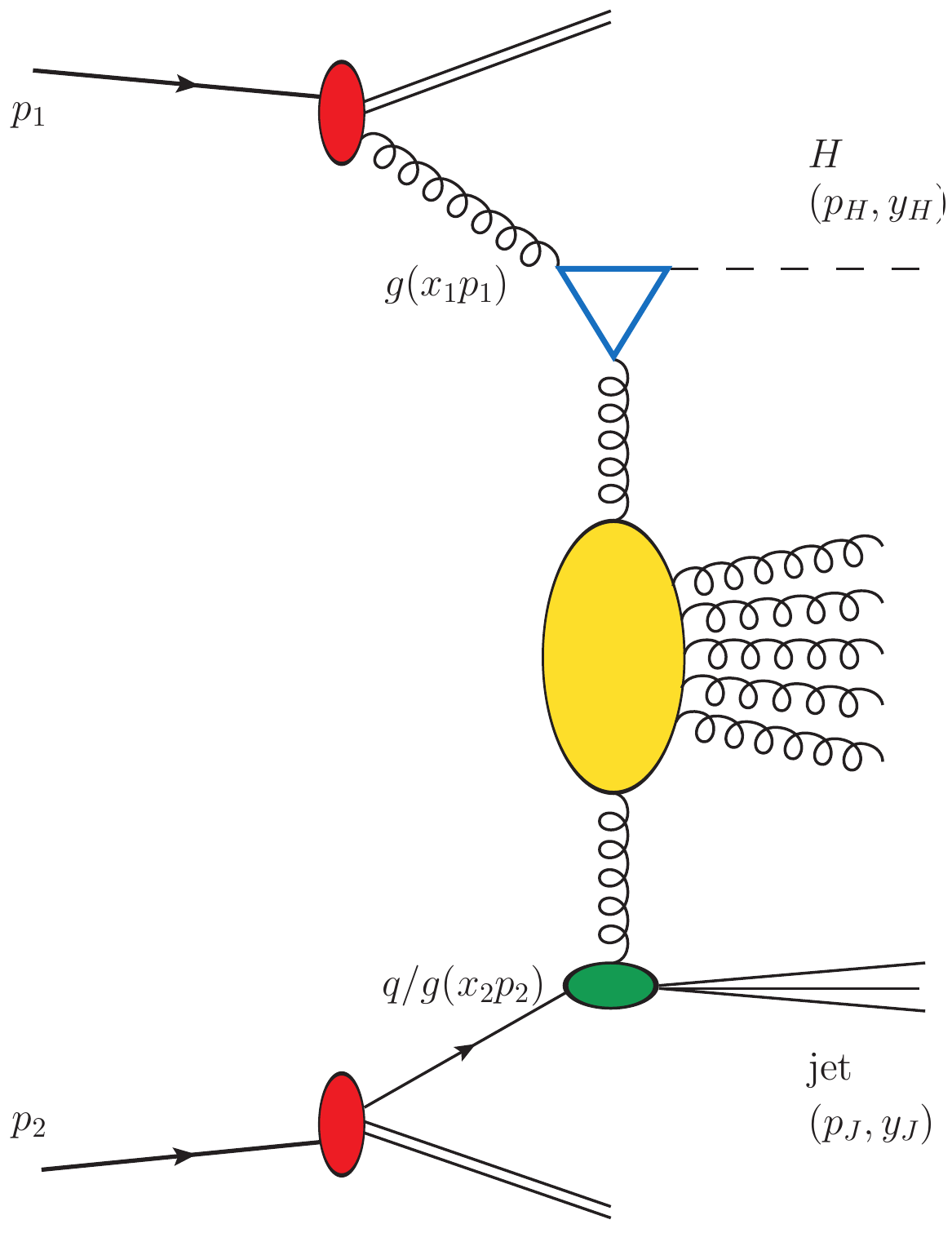}
\caption{Pictorial diagrammatic representation of the semi-hard Higgs-plus-jet hadroproduction. Taken from Ref.~\cite{Celiberto:2020tmb}.}
\label{fig:process}
\end{figure}

\begin{figure*}[!t]
\centering

   \includegraphics[scale=0.41,clip]{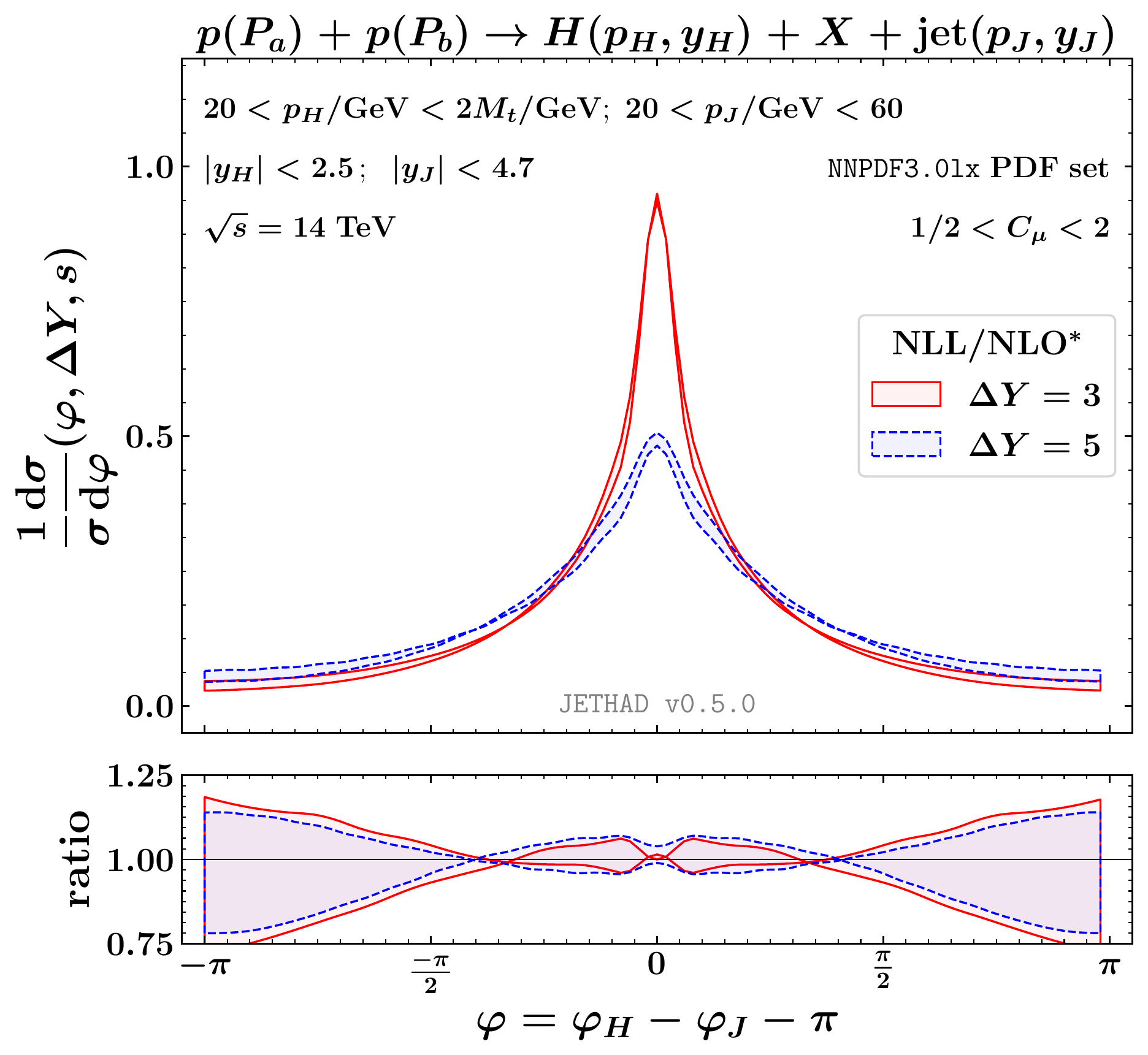}
   \hspace{0.10cm}
   \includegraphics[scale=0.41,clip]{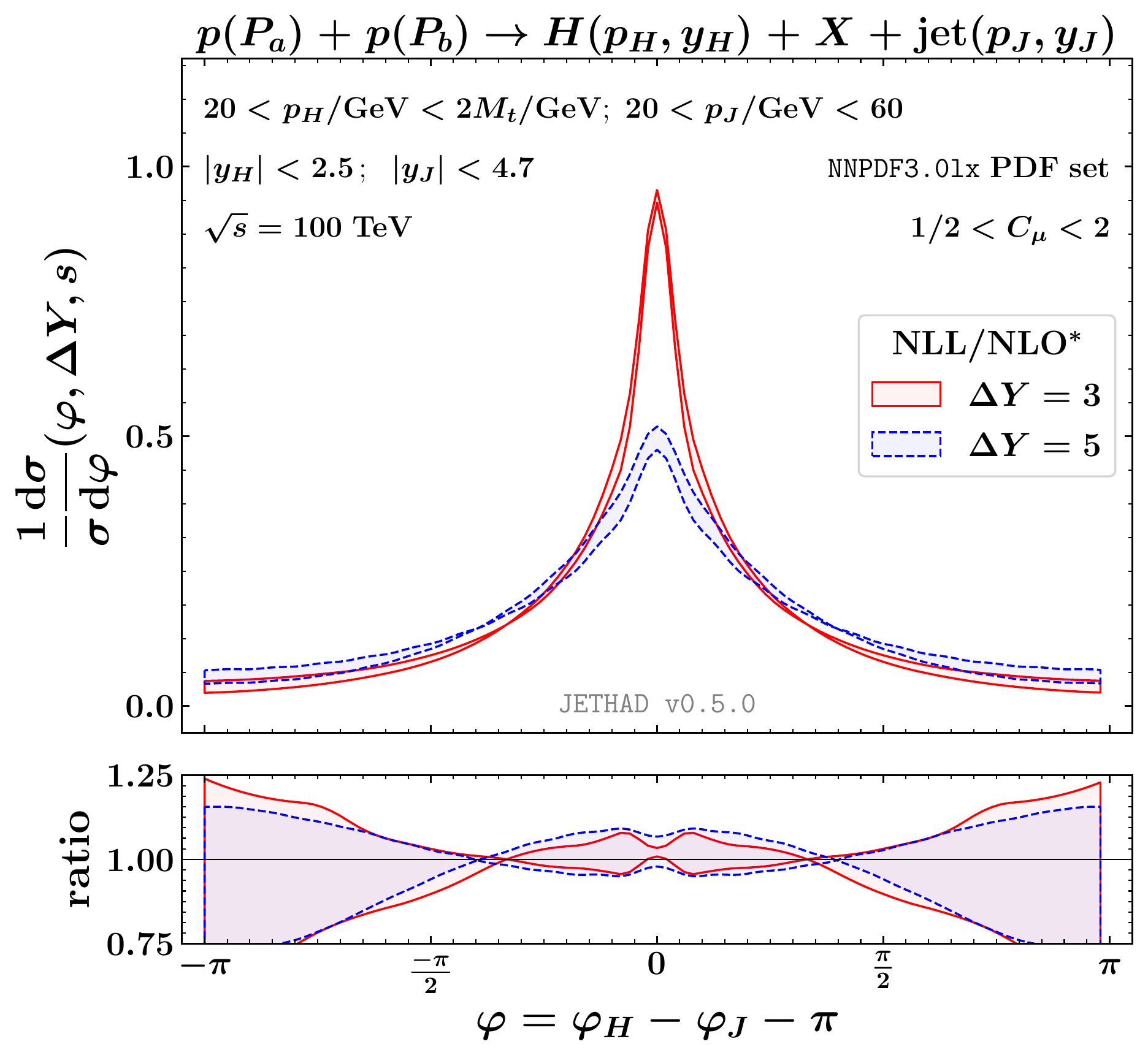}

\caption{Azimuthal-angle multiplicity for the Higgs-plus-jet production at the LHC (14~TeV, left) and the FCC (100~TeV, right).}
\label{fig:azimuthal_multiplicity}
\end{figure*}

\begin{figure*}[!t]
\centering

   \hspace{-0.30cm}
   \includegraphics[scale=0.41,clip]{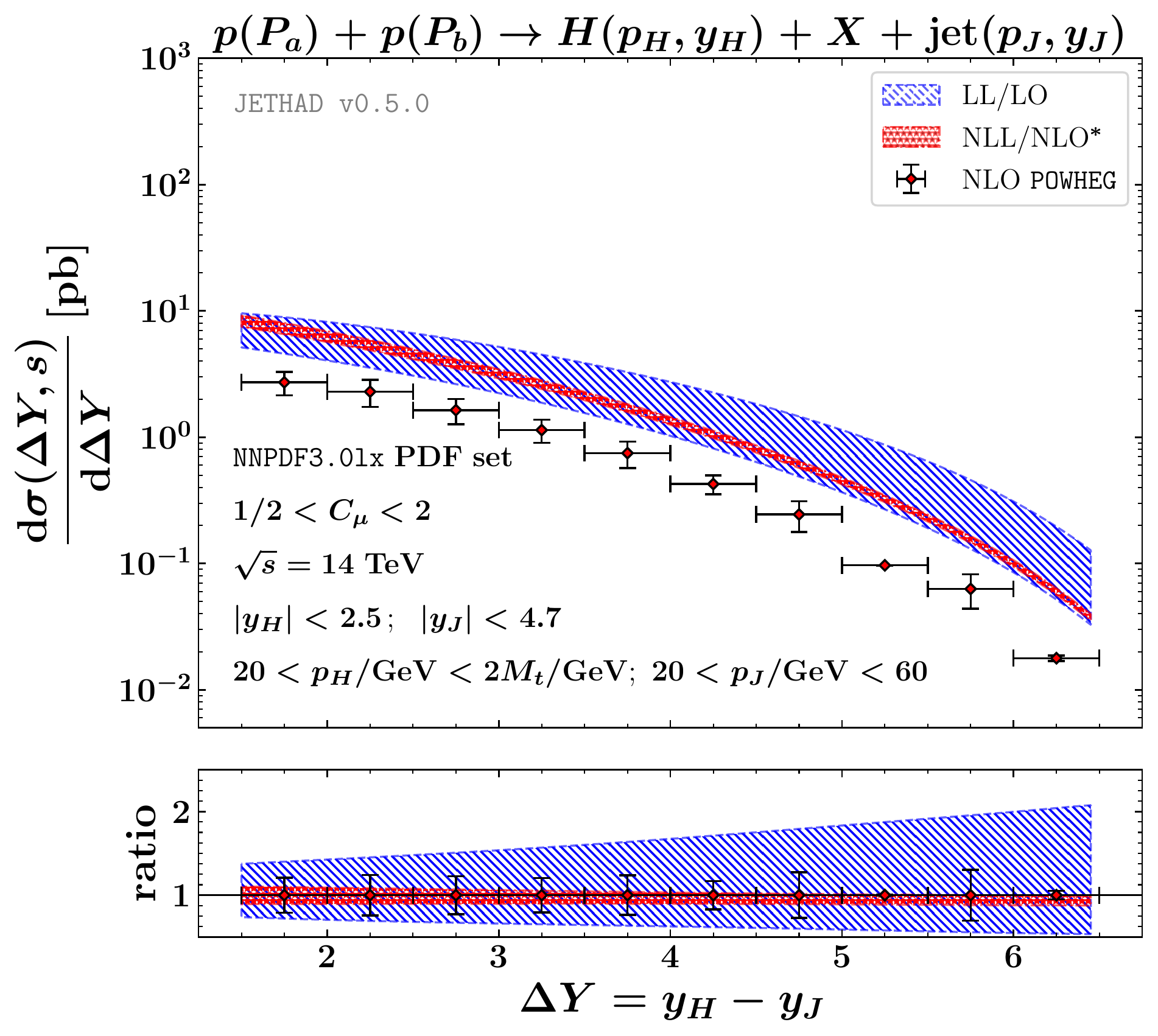}
   \hspace{-0.00cm} 
   \includegraphics[scale=0.41,clip]{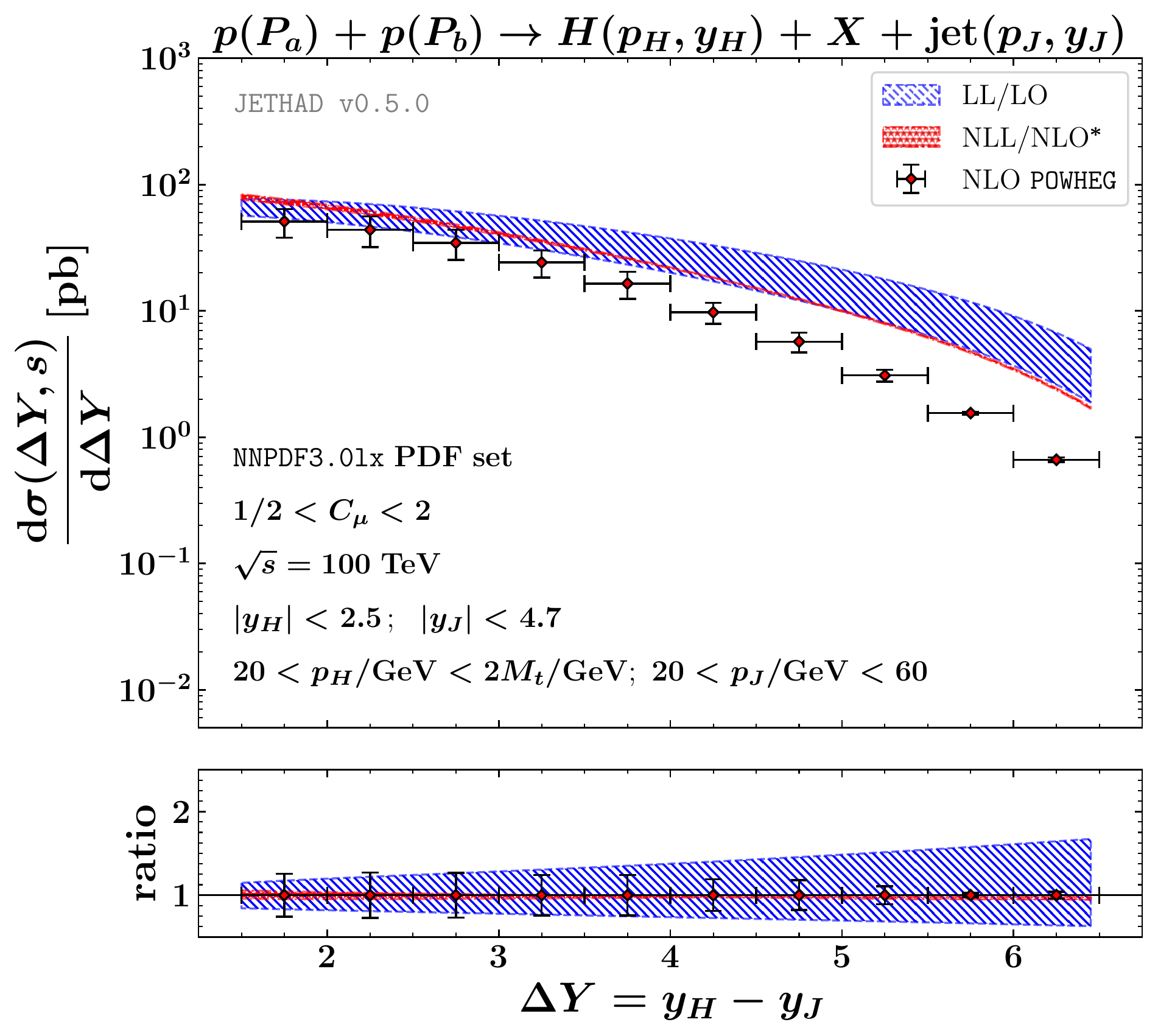}

\caption{Rapidity distribution for the Higgs-plus-jet production at the LHC (14~TeV, left) and the FCC (100~TeV, right).}
\label{fig:rapidity_distribution}
\end{figure*}

The reaction under investigation is (see Fig.~\ref{fig:process})
\begin{eqnarray}
\label{process}
 p(P_a) \!+\! p(P_b) \to H(p_H, y_H, \varphi_H) +\! X \!+\! {\rm jet}(p_J, y_J, \varphi_J) \, ,
\end{eqnarray}
where a Higgs boson with transverse-momentum modulus $p_H$, rapidity $y_H$, and azimuthal angle $\varphi_H$ is emitted in association with a light jet with transverse-momentum modulus $p_J$, rapidity $y_J$, and azimuthal angle $\varphi_J$, and together with a secondary, undetected gluon radiation, inclusively labeled as $X$. Both the Higgs boson and the jet have transverse masses which stay well above the QCD hadronization scale, $M_{\perp H,J} \gg \Lambda_{\rm QCD}$, with $M_{\perp H} = \sqrt{M_H^2 + p_H^2}$, $M_H = 125.18$~GeV, and $M_{\perp J} \equiv p_J$. Furthermore, they are separated by a large rapidity interval, $\DY = y_H - y_J$. 

It is convenient to rewrite the differential cross section as a Fourier series of azimuthal-angle coefficients, ${\cal C}_{n \ge 0}$
\begin{eqnarray}
 \label{dsigma_Fourier}
 \frac{\drv \sigma}{\drv p_H \drv p_J \drv y_{H,J} \drv \varphi} =
 \frac{1}{2\pi} \left[{\cal C}_0 + 2 \sum_{k=1}^\infty \cos (k \varphi)\,
 {\cal C}_n \right]\, .
\end{eqnarray}
Then, by relying upon the $\MSb$ renormalization scheme~\cite{PhysRevD.18.3998}, it is possible to write elegant and compact expressions for the ${\cal C}_n$ coefficients in the hybrid high-energy and collinear factorization (technical details are given in Section 2 of Ref.~\cite{Celiberto:2020tmb}).
Their validity holds up to the NLL/NLO$^*$ accuracy level. Indeed, NLO corrections to the Higgs singly off-shell coefficient function were calculated only recently~\cite{Celiberto:2022fgx,Hentschinski:2020tbi} and they have not yet been implemented in {\Jethad}, our reference technology for phenomenological studies~\cite{Celiberto:2020wpk,Celiberto:2022rfj}.

A comprehensive high-energy versus fixed-order analysis relies on confronting NLL resummed predictions with pure fixed-order calculations. Thus, we compare NLL/NLO$^*$ predictions for a selection of differential distributions, as implemented in {\Jethad}, with corresponding NLO fixed-order ones from {\Powheg}~\cite{Nason:2004rx,Frixione:2007vw,Alioli:2010xd,Campbell:2012am,Hamilton:2012rf}.
We stress that, in the present work, the {\Powheg} technology is used as a reference for purely fixed-order predictions.
We postpone to the future a relevant and interesting comparison between our resummed results and those obtained by considering transverse-momentum resummation effects through a \emph{parton-shower} approach (see, \emph{e.g.}, Refs.~\cite{Buckley:2021gfw,vanBeekveld:2022zhl,vanBeekveld:2022ukn} and references therein).
A two-loop QCD running coupling setup with a dynamic number of flavors is chosen. All calculations are performed in the $\MSb$ scheme.
We have also made preliminary analyses on gauging the effect of selecting different collinear PDF determinations~\cite{Harland-Lang:2014zoa,Dulat:2015mca,Butterworth:2015oua,NNPDF:2014otw,NNPDF:2021njg}, including sets obtained by resummation enhancements~\cite{Ball:2017otu,Bonvini:2015ira}.
A comprehensive study on uncertainties coming from PDFs is postponed to a future work. Here we make use of the large-$x$ improved {\tt NNPDF3.0lx} NLO PDF determination~\cite{Bonvini:2015ira}.

The Higgs boson is detected in rapidity acceptances typical of by the LHC barrels, 
$|y_H| < 2.5$, while the light jet also by LHC endcaps, $|y_J| < 4.7$, as in Ref.~\cite{Khachatryan:2016udy}.
The Higgs transverse momentum ranges from 20~GeV to two times the top-quark mass, $M_t$, whereas the jet one stands from 20 to 60~GeV.
We consider two values for the center-of-mass energy: is $\sqrt{s} = 14$~TeV~(LHC-like) and $\sqrt{s} = 100$~TeV~(FCC-nominal).
Shaded bands in our plots refer to the effect of varying the renormalization and the factorization scale from $1/2$ to two times the \emph{natural} values suggested by final-state kinematics.

The first observable under investigation is the \emph{azimuthal-angle multiplicity}~\cite{Marquet:2007xx,Ducloue:2013hia,Celiberto:2020wpk}, namely a normalized cross section differential in and $\DY$ and $\varphi$
\begin{eqnarray}
 \label{azimuthal_multiplicity}
 \frac{1}{\sigma} \frac{\drv \sigma}{\drv \DY \, \drv \varphi} = \frac{1}{2 \pi} \left[ 1 + 2 \sum_{k=1}^\infty
 \cos (k \varphi) \, R_{k0} \right] \, ,
\end{eqnarray}
where $R_{k0} \equiv C_k/C_0$ are the so-called azimuthal-cor\-re\-lat\-ion moments, with $C_k$ the azimuthal coefficients in Eq.~\eqref{dsigma_Fourier} integrated over the previously given $y_{H,J}$ and $p_{H,J}$ cuts, while the rapidity interval, $\Delta Y = y_H - y_J$, is kept fixed.
The azimuthal multiplicity actually represents one of the most favorable observables whereby hunting for high-energy imprints. Indeed, it collects the whole high-energy signal coming from all azimuthal modes, and not from a single azimuthal correlation. Furthermore, since experimentally measured distributions generally do not cover the whole $(2 \pi)$ range of azimuthal angle due to limitations of the detector, observables differential on $\varphi$ allow us to quench the accuracy loss. Finally, as proven in Ref.~\cite{Celiberto:2022kxx}, azimuthal distributions  permit us to partially overcome the well-known instabilities affecting the description of light-flavor sensitive final states.

Results for the azimuthal multiplicity at NLL/NLO$^*$ are presented in Fig.~\ref{fig:azimuthal_multiplicity} at the LHC (left) and the FCC (right).
Ancillary panels below primary plots show reduced distributions, divided by their central value calculated at natural scales.
From the inspection of these results, the onset of high-energy dynamics fairly emerges.
As a general pattern, one observes the presence of a narrow peak at $\varphi = 0$, namely when the Higgs boson and the light jet are emitted back-to-back.
Both at LHC and FCC energies the peak height lowers very fast as $\DY$ grows, while the two distribution tails slightly widen.
As predicted by the high-energy resummation, the weight of the secondary gluon emissions, produced with a strong hierarchy in rapidity, increases with $\DY$. This brings to an increase of the Higgs-jet decorrelation in the azimuthal plane, so that the amount of back-to-back events falls off.
Here, possible small transverse-momentum imbalance contaminations could be relevant and they the should be resummed as well~\cite{Mueller:2012uf,Marzani:2015oyb,Mueller:2015ael,Xiao:2018esv,Hatta:2021jcd,Hatta:2020bgy}.

The second observable matter of our study is the \emph{rapidity distribution}, namely the cross section differential in $\DY$. It genuinely corresponds to the first azimuthal coefficient, $C_0$, taken from Eq.~\eqref{dsigma_Fourier} and integrated over the previously given $y_{H,J}$ and $p_{H,J}$ ranges.
NLL/NLO$^{*}$ results for the $\DY$-distribution are compared with the LL/LO limit and with NLO fixed-order calculations from {\Powheg}, see left (right) plot of Fig.~\ref{fig:rapidity_distribution} for the corresponding analysis at LHC (FCC) energies.
The information contained in main plots is complemented by the one about reduced cross sections in ancillary panels.
Here, the usual onset of the high-energy dynamics easily comes out. The growth with energy of the partonic hard factors is dampened by the convolution with PDFs. 
This globally generates a falloff of physical distributions as the rapidity distance grows.
We observe that NLL resummed bands (red) are almost completely nested inside pure LL ones (blue) for both energies considered. This further corroborates the assumption, made in Ref.~\cite{Celiberto:2020tmb}, that the large energy scales connected to Higgs-boson transverse masses act as fair \emph{stabilizers} of the high-energy series.

The large discrepancy between resummed predictions \emph{\`a la} {\Jethad} and fixed-order results \emph{via} {\Powheg} in Fig.~\ref{fig:rapidity_distribution} motivates future developments focused on matching the two approaches.
We believe that such a matching procedure is a required step to assess the actual weight of NLL high-energy corrections on top of next-to-leading QCD calculations of Higgs-plus-jet distributions from LHC to FCC collision energies.

\section{Towards NLL matched to NLO}
\label{sec:conclusions}

We presented arguments supporting the emergence of a \emph{natural stabilization} of the high-energy resummation at NLL \emph{via} the semi-inclusive detection of a Higgs bosons in association with a light-flavored jet at hadron colliders. 
By relying on a NLL/NLO$^*$ hybrid high-energy and collinear treatment for rapidity and azimuthal-angle distributions, we hunted for high-energy signals from LHC~(14~TeV) to FCC~(100~TeV) center-of-mass energies.
These signals are present and motivate future analyses aimed at gauging the feasibility of precision studies of
QCD in its high-energy limit \emph{via} our hybrid factorization.
Future steps are: $(i)$ the implementation in {\Jethad} of the full NLO correction to the forward-Higgs off-shell coefficient function~\cite{Celiberto:2022fgx,Hentschinski:2020tbi}, $(ii)$ the definition of a matching procedure between NLL resummed predictions and NLO fixed-order calculations~\cite{Celiberto:2023uuk}, and $(iii)$ assessing the weight of heavy-quark finite-mass corrections~\cite{Bonciani:2016qxi,Jones:2018hbb}, including also bottom flavor~\cite{Bonciani:2022jmb}, which could be potentially relevant at FCC collision energies.

\vspace{-0.10cm}

\section*{Acknowledgements}

This work was supported by the Atracci\'on de Talento Grant n. 2022-T1/TIC-24176 of the Comunidad Aut\'onoma de Madrid, Spain, and by the INFN/NINPHA and INFN/QFT @COLLIDERS Projects, Italy.
F.G.C. thanks the Universit\`a degli Studi di Pavia for the warm hospitality.

\vspace{-0.10cm}

\bibliographystyle{elsarticle-num}

\bibliography{bibliography}

\end{document}